\documentclass[12pt]{article}
\usepackage{graphicx, subfigure}
\usepackage{amssymb}
\usepackage{amscd}
\usepackage{amsmath}
\usepackage{appendix}
\usepackage{verbatim}

\begin{document}
\begin{titlepage}

{\hbox to\hsize{\hfill June 2016 }}

\bigskip \vspace{3\baselineskip}

\begin{center}
{\bf \Large

Heavy Leptonium as the Origin of the 750 GeV Diphoton Excess }

\bigskip

\bigskip

{\bf  Neil D. Barrie$^1$, Archil Kobakhidze$^1$, Shelley Liang$^1$, Matthew Talia$^1$ and Lei Wu$^{1,2}$ \\ }

\smallskip

{ \small \it
$^1$ ARC Centre of Excellence for Particle Physics at the Terascale, \\
School of Physics, The University of Sydney, NSW 2006, Australia \\
$^2$ Department of Physics and Institute of Theoretical Physics, Nanjing Normal University, \\ Nanjing, Jiangsu 210023, China\\
E-mails:  neil.barrie, archil.kobakhidze, shelley.liang, matthew.talia, lei.wu1@sydney.edu.au
\\}

\bigskip

\bigskip

\bigskip

{\large \bf Abstract}

\end{center}
\noindent We propose a hypothetical heavy leptonium, the scalar bound state of an exotic lepton-antilepton pair, as a candidate for the recent 750 GeV resonance in the early LHC Run 2 data. Such a para-leptonium is dominantly produced via photon-photon fusion at the LHC and decays into a photon pair with a significant branching fraction. In addition, our model predicts a companion spin-1 ortho-leptonium bound state, which can decay to $W^+W^-$, $f\bar{f}$ and three photons. Under the LHC and the electroweak precision observables bounds, we find that the observed excess of 750 GeV diphoton events can be explained within $2\sigma$ accuracy for $Y_{L} \approx 4.8 - 7.2$ for the minimal case in our scenario. The observation of the ortho-leptonium in the dilepton channel will be the smoking gun for our scenario.
\end{titlepage}

\section{Introduction}

The ATLAS and CMS Collaborations have recently reported excesses of two-photon events in 13 TeV proton-proton collisions at the LHC at $3.9\sigma$ ($2.3\sigma$ global) \cite{ATLAS750} and $2.6\sigma$ ($1.2\sigma$ global) \cite{CMS750} local significance, respectively. The number of observed events, if interpreted as the production of a 750 GeV resonance, requires the following cross sections for the process  $pp\to R\to \gamma\gamma$ \cite{Franceschini:2015kwy}:
\begin{eqnarray}
\label{1.1}
\sigma^{\rm ATLAS}(pp\to R \to \gamma\gamma)&\approx &(10\pm 3)~ {\rm fb} \\
\label{1.2}
\sigma^{\rm CMS}(pp\to R \to \gamma\gamma)&\approx & (6\pm 3)~ {\rm fb}
\end{eqnarray}
Combining ATLAS and CMS data and including 8 TeV data lead to the estimate \cite{Buttazzo:2015txu}:
\begin{equation}
\sigma (pp\to R \to \gamma\gamma) \approx  (4.6\pm1.2)~ {\rm fb}
\label{comb}
\end{equation}
These initial reports have been confirmed by the ATLAS and CMS Collaborations at the recent Moriond conference, with a slightly increased significance \cite{atlas_moriond, cms_moriond}. This further motivates us to explore different theoretical models that incorporate the corresponding resonance $R$ without conflicting with other available experimental data.

The reported excess seems to show up only in the diphoton channel, without accompanying leptons or jets or any significant missing energy.
This implies a relatively large coupling of the resonance to photons, and hence large branching fraction ${\rm Br}(R\to\gamma\gamma)$, and may hint that the putative resonance is produced predominantly via photon fusion \cite{Fichet:2015vvy,Csaki:2015vek,Harland-Lang:2016qjy}. There are only a few models so far that make use of photoproduction as a dominant production mechanism for the 750 GeV resonance. The desired large $R\gamma\gamma$ coupling has been obtained in an extra-dimensional model of Ref. \cite{Abel:2016pyc} via loops of a Kaluza-Klein  tower of charged leptons. In the composite axion scenario of Ref. \cite{Barrie:2016ntq} the enhancement of $R\gamma\gamma$ coupling has been achieved due to the higher-colour representation of hypothetical vector-like quarks. The stringy model has been suggested in \cite{Anchordoqui:2015jxc} and a hypercharge axion model is proposed in \cite{Ben-Dayan:2016gxw}.

In this paper we suggest an economical extension of the Standard Model by introducing a vectorlike, weak isospin singlet lepton $L$ with mass $m_{L}$ and hypercharge (equal to the electric charge) $Y_{L}$. Such leptons are typically stable and form an unstable heavy spin-0 para-leptonium state $\psi_{para} \sim \bar L L$, which can be identified as the recent 750 GeV diphoton resonance\footnote{Exotic heavy quarkonia have also been suggested as an explanation of the LHC diphoton excess \cite{Luo:2015yio, Han:2016pab, Kats:2016kuz, Kamenik:2016izk,Bi:2016gca}.}. In what follows we compute the leptonium production rate and constrain the parameter $Y_{L}$ to fit the diphoton data \cite{ATLAS750,CMS750}, simultaneously satisfying the electroweak precision observables and the LHC constraints.

\section{Heavy Leptoniums and Constraints}
In the non-relativistic approximation, the heavy lepton $L$ is described by the Schr\"odinger equation
\begin{equation}
\left(-\frac{\nabla^2}{m_{L}} + V(r)\right)\psi = E\psi ~,
\label{2.1}
\end{equation}
with the binding Coulomb potential
\begin{equation}
V(r)=-\frac{Y_{L}^{2}\alpha}{r}~,
\label{2.2}
\end{equation}
where $\alpha \approx 1/128$ is the fine structure constant evaluated at $m_Z$ \footnote{Note that since the leptonium Bohr radius, $r_{L}=\frac{2}{m_{L}}\frac{1}{Y_{L}^{2}\alpha}$, is larger than the Compton wavelength of the $Z$-boson, $r_{L} \gtrsim 1/m_{Z}$, in deriving the potential (\ref{2.2}) we only take into account massless photon exchange.}. In this approximation, the ground state ($n=1,~ l=0$) energy is given by:
\begin{equation}
E=-\frac{1}{4}m_{L}\left(Y_{L}^2\alpha\right)^{2}~,
\label{2.3}
\end{equation}
We also include the leading $\left[ \sim \mathcal{O}\left(Y_{L}^8\alpha^4\right)\right]$ relativistic Breit correction to the binding energy given in Eq. (\ref{2.3}):
\begin{equation}
\delta E_{\rm Breit}=-\frac{1}{2m_{L}}\left(E^{2}-2E\left\langle V\right\rangle +\left\langle V^{2}\right\rangle \right)=-\frac{5}{16}m_{L}\left(Y_{L}^{2}\alpha\right)^{4}~.
\label{2.4}
\end{equation}
The mass of the para-leptonium $\psi_{para}$ ($J^{PC}=0^{-+}$) is then given by:
\begin{equation}
m_{\psi_{para}}=2m_{L}+E+\delta E_{\rm Breit},
\label{2.5}
\end{equation}
and from this we can derive the mass of the lepton, if it is assumed that the 750 GeV resonance can be identified as the para-leptonium state,
\begin{equation}
	m_L=m_{\psi_{para}}\left(2-\frac{1}{4}(\alpha Y_{L}^2)^2-\frac{5}{16}(\alpha Y_{L}^2)^4\right)^{-1}~.
	\label{lepton_mass}
\end{equation}

The wave function $\psi_{para}(0)$ is given by,
\begin{equation}
\left|\psi_{para}(0)\right|^{2}=\left(\frac{1}{\sqrt{4\pi}}\right)^2\left|R_{para}(0)\right|^{2}
\label{2.8}
\end{equation}
with the radial part evaluated as:
\begin{equation}
\left|R_{para}(0)\right|^{2}=\frac{\left(Y_{L}^{2}\alpha m_{L}\right)^{3}}{2}
\label{2.8b}
\end{equation}

\begin{table}[!htb]
\begin{center}
\begin{tabular}{c|cccccc}
\hline\hline
Mass (GeV) &~~ ~~100~~ & ~~200~~ & ~~300~~ & ~~400~~ & ~~450~~ & ~~500~~\\
\hline
Electric charge (e) & ~~~~$1-5$~~ & ~~$1-5$~~ & ~~$1-5$~~ & ~~$1-5$~~ & ~~$1.3-5$~~ & ~~$2.2-5$~~\\
\hline\hline
\end{tabular}
\end{center}
\caption{CMS limit on multi-charged stable particles \cite{cms-multicharged}.}
\label{table:multi-charged}
\end{table}
These results are obtained through expansion around powers of $ \alpha Y_L^2 $. For the range of hypercharges considered in this analysis, the calculations undertaken are done so well within the perturbative regime.
In the following study, we will assume that the para-leptonium $\psi_{para}$ is the observed 750 GeV diphoton resonance.

The ATLAS and CMS collaborations have searched for multi-charged heavy stable particles \cite{Chatrchyan:2013oca, atlas-multicharged}. The latest constraints are imposed on particles with integer charge (in units of electron charge $e$), $q=ze,~z\in \mathbb{Z}$, which require that they are heavier than $\sim 700 - 900$ GeV, assuming Drell-Yan production. Note, however, that these bounds are not directly applicable to non-integer charged particles. The earlier CMS bounds \cite{cms-multicharged} on multi-charged particles in the range $1-5e$ are given in Tab.~\ref{table:multi-charged}. It should also be noted, that the LHC analysis does not take into account the possibility of the production of bound states instead  of pairs of free particles, which may significantly alter the searches. On the other hand, the abundance of fractionally charged particles is constrained to be less than $\sim 10^{-22}$ per ordinary matter nucleon \cite{Perl:2004qc} from the Millikan-type experiments. This bound can be easily satisfied by assuming a low reheating temperature after inflation, $T_r < m_{L}$. Alternatively, for certain fractional charge assignments our leptons can decay via high-dimensional operators fast enough to result in a sufficiently low abundance. The decaying exotic leptons will also evade collider constraints.

\begin{figure}[ht]
	\centering
\includegraphics[width=80mm]{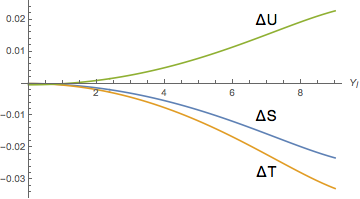}
	\caption{\small The exotic lepton contributions to the oblique parameters $ \Delta  S $, $ \Delta  T $, $ \Delta  U $ as a function of the lepton hypercharge.}
	\label{fig:3}
\end{figure}

Since the multi-charged lepton can contribute to the electroweak observables, we present the dependence of the $\Delta S,\Delta T,\Delta U$ parameters \cite{Peskin:1991sw} on the hypercharge of the new lepton in Fig.~\ref{fig:3}. We found that our predictions of the oblique parameters are consistent with the experimental measurements \cite{pdg}.

In addition to para-leptonium, our model predicts a heavier spin-1 ortho-leptonium bound state $\psi_{ortho}$ ($J^{PC}=1^{--}$) with mass  \cite{Efimov:2010ih},
\begin{equation}
m_{\psi_{ortho}}\simeq m_{\psi_{para}}\left(1 + \frac{7}{12}\frac{\left(Y_{L}^{2}\alpha\right)^{4}}{\left(2-\frac{1}{4}\left(Y_{L}^{2}\alpha\right)^{2}-\frac{5}{16}\left(Y_{L}^{2}\alpha\right)^{4}\right)}\right),
\label{ortho}
\end{equation}
Such a spin-1 bound state can be singly produced via quark-antiquark annihilation at the LHC. At leading order, the Drell-Yan cross section of $q\bar{q} \to \psi_{ortho}$ is given by \cite{Schuler:1994hy}
\begin{eqnarray}
	\sigma_{q\bar{q} \to \psi_{ortho}} =
	3\sum_{q=u,d,s}
	\frac{\Gamma(\psi_{ortho} \to q\bar{q})}{s\,m_{\psi_{ortho}} }
	\left[ \frac{4\pi^2}{9} \int dx_1 dx_2 \delta(x_1 x_2 - m_{\psi_{ortho}}^2/s)\,f_q(x_1) f_{\bar{q}}(x_2) \right],
\end{eqnarray}
where $f_q(x)$ and $f_{\bar{q}}(x)$ denote the quark and antiquark PDFs. It will decay into $W^+W^-$, SM fermion pairs $f\bar{f}$ ($f=e,\mu,\tau,u,d,c,s,b,t$) via an intermediate photon \cite{Barger:1987xg,Hewett:1988ay}, or a three photon final state \cite{3photon}. In the limit $m_{\psi_{ortho}} \gg m_W, m_f$, the main partial decay widths are given by,
\begin{eqnarray}
\Gamma(\psi_{ortho} \to W^+W^-) &=& \frac{\alpha^2 Y^2_L m^2_{\psi_{ortho}}}{16 m^4_W} |\psi_{ortho}(0)|^2, \\
\Gamma(\psi_{ortho} \to f\bar{f}) &=& \frac{16 \pi \alpha^2 N_c }{m^2_{\psi_{ortho}}} Y^2_L Q^2_f |\psi_{ortho}(0)|^2, \\
\Gamma(\psi_{ortho} \to 3\gamma) &=& \frac{64\alpha^3 }{9m^2_{\psi_{ortho}}}Y^6_L |\psi_{ortho}(0)|^2,
\end{eqnarray}
where $N_c$ is a colour factor (1 for leptons, 3 for quarks). The wave function $\psi_{ortho}(0)$ is equal to $\psi_{para}(0)$ at leading order. Among these decay channels, the LHC searches for dilepton final states will give the most stringent limit, as shown in Tab.~\ref{ortho}, which requires that the heavy lepton carries a hypercharge $Y_L\lesssim 5.6(5.2)$ by ATLAS(CMS) at 13 TeV LHC. It should be noted that this is in the minimal scenario, when the ortho-leptonium only decays into standard model particles. Below we allow the total width of the para-leptonium to vary, so this bound will only be applicable to the minimal case and thus would be loosened accordingly.
\begin{table}[h]
	\centering
	\caption{The upper bounds on $\sigma\left(pp\to Z^\prime\right){\text{Br}} \left(Z^\prime\to\ell\ell\right)$ from the 13 TeV LHC. The units used in this table are fb.}
	\begin{tabular}{l|r|r|rrrrrrr}
    \hline\hline
	\quad\quad @13TeV & ATLAS \cite{atlas-2l} & CMS \cite{cms-2l} & $Y_L=4.7$  & 4.8 & 4.9 & 5 & 5.1 & 5.2 & 5.6\\
	\hline
	$\quad\quad\, \psi_{ortho} \to L^+ L^-$ & 5.6 & 2.9 & 1.35  & 1.61  & 1.89  & 2.22 & 2.60 & 3.03 & 5.44\\
	\hline\hline
	\end{tabular}
	\label{ortho}
\end{table}


\section{Photoproduction of 750 GeV Para-leptonium}
\begin{figure}[ht]
	\centering
    \includegraphics[width=\textwidth]{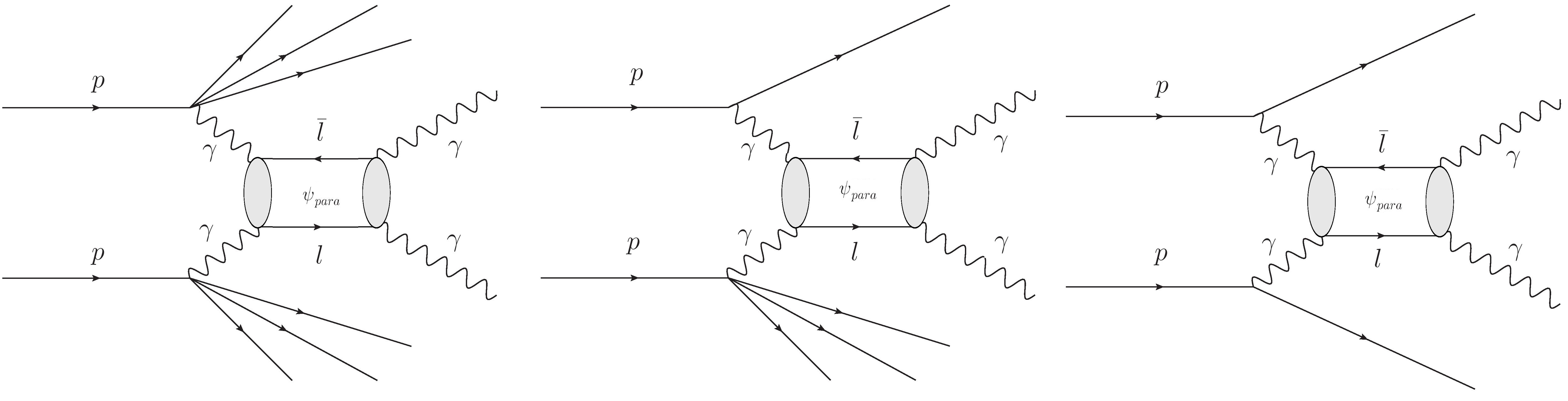}
	\caption{\small The Feynman diagrams for inelastic, semi-elastic and elastic scattering
photoproduction subprocesses, from left to right .}
	\label{feyn}
\end{figure}

At the LHC the spin-0 para-leptonium is predominantly produced via photon-fusion processes \footnote{Recently, in \cite{Ababekri:2016kkj}, it has been shown that the tension of photoproduction of a 750 GeV scalar in 8 TeV and 13 TeV LHC data can be relaxed due to improvements of the photon PDFs.} in three distinct ways, shown in Fig.~\ref{feyn}. Photoproduction is dominated by inelastic scattering, which is followed by the semi-elastic and elastic processes in the ratio 63:33:4 \cite{Csaki:2015vek}. The parton level cross section for photoproduction of the bound state $\psi_{para}$ can be written in terms of the decay width of $\psi_{para}\to\gamma\gamma$, since both the production and decay processes share the same matrix elements. This is given by \cite{Bertulani:2001zk, Kats:2012ym}:
\begin{equation}
    \hat{\sigma}_{\gamma \gamma \rightarrow\psi_{para}}(\hat{s})=8 \pi^2\frac{\Gamma_{\psi_{para} \rightarrow \gamma\gamma}}{m_{\psi_{para}}} \delta(\hat{s}-m^2_{\psi_{para}})
\label{2.7}
\end{equation}
The decay width $\Gamma_{\psi_{para} \rightarrow \gamma\gamma}$ in turn is given in terms of the annihilation cross section of a free lepton-antilepton pair into two photons and the  wave function for the leptonium bound state evaluated at the origin:
\begin{equation}
    \Gamma_{\psi_{para} \rightarrow \gamma \gamma}=\frac{16\pi\alpha^2 Y^4_{L}\left|\psi_{para}(0)\right|^{2}}{m^2_{\psi_{para}}}~.
\end{equation}
Then we can calculate the two photon production and decay cross section by convolution with the parton distribution function (PDF) for the photon in the proton, $f_{\gamma}(x)$:
\begin{equation}
    \sigma_{\gamma\gamma\rightarrow\psi_{para}\rightarrow\gamma\gamma}=\frac{8 \pi^2}{sm_{\psi_{para}}}\frac{\Gamma^2_{\psi_{para} \rightarrow \gamma\gamma}}{\Gamma_{\psi_{para}}} \int\delta(x_{1}x_{2}-m^2_{\psi_{para}}/s)f_{\gamma}(x_1)f_{\gamma}(x_2)dx_{1}dx_{2}
 \label{2.9}
\end{equation}
where we have set the centre-of-mass energy to $\sqrt{s}=13$ TeV and $m_{\psi_{para}}=750$ GeV. $\Gamma_{\psi_{para}}$ in the above equation denotes the leptonium total decay width. In the numerical calculations, we assume the total decay width of the 750 GeV para-leptonium as a free parameter but with an upper bound $\Gamma_{\psi_{para}}\leq 45$ GeV  motivated by the ATLAS measurement \cite{ATLAS750}. We use the \texttt{NNPDF2.3QED} PDF \cite{Ball2013} to calculate the hadronic cross section at the LHC.

The potentially large decay width, $\Gamma_{\psi_{para}}\leq 45$ GeV, can be realized by introducing additional light particles $X$ (e.g. dark matter particles) to couple with para-leptonium. Such an extension can relax the LHC constraints on multi-charged heavy stable particles in Tab.\ref{table:multi-charged} and also be consistent with other LHC data when the mass splitting between $X$ and $L$ is small \cite{Han:2016pab}. The constraints imposed on this model by the ortho-leptonium bounds can also be relaxed in this way, as the derived bounds assume the minimal total decay width; only to SM particles. 

\begin{figure}[t]
	\centering
\subfigure{\label{fig:crosssec1}\includegraphics[width=90mm,height=90mm]{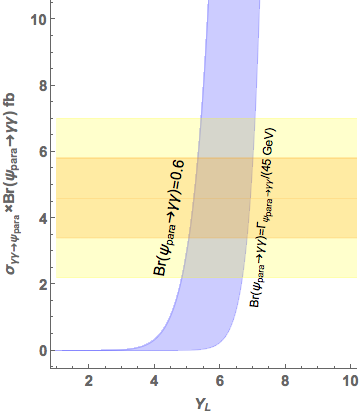}}\hspace{5mm}
\caption{\small The photoproduction cross section of para-leptonium $\psi_{para}$ at 13 TeV LHC. The boundaries of the curves from left to right correspond to a hypercharge-coupling only width to a total width of 45 GeV. The $1\sigma$ (orange/dark) and $2\sigma$ (yellow/light) range of the diphoton excess of the combined result in Eq.~(\ref{comb}) are shown.}
\label{fig:1}
\end{figure}
In Fig.~\ref{fig:1}, we present the dependence of the photoproduction rate of the process $pp \to \psi_{para} \to \gamma\gamma$, on the hypercharge quantum number $Y_{L}$ at 13 TeV LHC. The $1\sigma$ (orange/dark) and $2\sigma$ (yellow/light) range of the diphoton excess of the combined result in Eq.~(\ref{comb}) are shown. From Fig.~\ref{fig:1}, we can see that the hypercharge is required to be in the range $Y_{L} \approx 4.8 - 7.2$ to explain the LHC diphoton excess within the $2\sigma$ range.  For this allowed charge range the mass of the lepton is between $  \sim $ 375 and 385 GeV, from Eq. (\ref{lepton_mass}). For the minimal scenario, we find that the hypercharge range $Y_{L} \approx 4.8 - 5.4$ is favoured by the current LHC data on the diphoton excess. It is found that the ortho-leptonium constraint is consistent with this hypercharge range. In addition, if we allow the total width to be a free parameter, as discussed above, we expect that the corresponding ortho-leptonium bounds to apply no further constraints on the allowed hypercharge range obtained for larger total widths. It is expected that the future precision measurements of dilepton channel will further test our interpretation of the 750 GeV excess at the LHC.

\section{Conclusion}
In this paper, we have proposed a heavy spin-0 leptonium bound state of exotic vector-like leptons with high hypercharge $Y_{L}$ as a source of the LHC diphoton excess at 750 GeV invariant mass. The leptonium is predominantly produced in photon-fusion at the LHC and decays into two photons. We have found that for $Y_{L}\approx 4.8 - 7.2$ the cross section needed to explain the diphoton excess can be reproduced and is consistent with LHC bounds on the ortho-leptonium state. This model predicts a spin-1 bound state which decays to $W^+W^-$, $f\bar{f}$ and 3$\gamma$ final states. The observation of the ortho-leptonium in the dilepton channel will be the smoking gun for our scenario.

\paragraph{Acknowledgement.} This work was partially supported by the Australian Research Council. AK was also supported in part by the Shota Rustaveli National Science Foundation (grant DI/12/6- 200/13).

\end{document}